\documentclass{aastex}
\usepackage{emulateapj5}

\newcommand{\OVI}   {\ion{O}{6}}

\newcommand{\CIV}   {\ion{C}{4}}
\newcommand{\CIII}  {\ion{C}{3}}
\newcommand{\NV}    {\ion{N}{5}}
\newcommand{\NII}   {\ion{N}{2}}

\newcommand{\SII}   {\ion{S}{2}}
\newcommand{\FeIII} {\ion{Fe}{3}}
\newcommand{\FeII}  {\ion{Fe}{2}}
\newcommand{\SiIV}  {\ion{Si}{4}}
\newcommand{\SiIII} {\ion{Si}{3}}
\newcommand{\SiII}  {\ion{Si}{2}}
\newcommand{\HII}   {\ion{H}{2}}
\newcommand{\Ha}    {H$\alpha$}
\newcommand{\HI}    {\ion{H}{1}}
\newcommand{\HH}    {H$_2$}

\newcommand{\flux}  {erg~s$^{-1}$~cm$^{-2}$}
\newcommand{\fluxd} {erg~s$^{-1}$~cm$^{-2}$~\AA$^{-1}$}
\newcommand{\surfb} {erg~s$^{-1}$~cm$^{-2}$~arcsec$^{-2}$}
\newcommand{\kms}   {km~s$^{-1}$}
\newcommand{\dd}    {\arcdeg}
\newcommand{\SNR}   {SNR\,0057$-$7226}

\begin{document}

\title{Far-Ultraviolet \& \Ha\ Spectroscopy of SNR\,0057-7226 in the SMC \HII\ Region N66}
\author{
Charles W. Danforth,\altaffilmark{1,}\altaffilmark{3} 
Ravi Sankrit,\altaffilmark{1} 
William P. Blair,\altaffilmark{1} 
J. Christopher Howk,\altaffilmark{1,}\altaffilmark{4} \& 
You-Hua Chu,\altaffilmark{2,}\altaffilmark{3}
} 

\altaffiltext{1}{Department of Physics and Astronomy, The Johns Hopkins University, 3400 N. Charles Street,  Baltimore, MD 21218; danforth@pha.jhu.edu, ravi@pha.jhu.edu, wpb@pha.jhu.edu}
\altaffiltext{2}{Astronomy Department, University of Illinois, 1002 W. Green Street, Urbana, IL 61801; chu@astro.uiuc.edu}
\altaffiltext{3}{Visiting astronomer, Cerro Tololo Inter-American Observatory.}
\altaffiltext{4}{Current address: Center for Astrophysics \& Space Sciences, Univ. of California, San Diego, 9500 Gilman Dr., La Jolla, CA, 92093; howk@ucsd.edu }

\begin{abstract}
N66/NGC\,346 is the largest and brightest \HII\ region in the Small Magellanic Cloud and contains at least one known supernova remnant \SNR.  Optical emission from the remnant is overwhelmed by the bright photoionized emission from the nebula, but the remnant has been detected by way of far ultraviolet absorption lines.  Here we present data from the Far Ultraviolet Spectroscopic Explorer ({\it FUSE}) satellite showing strong \OVI\ and \CIII\ emission from a position at the edge of \SNR.  We also present high-resolution, long-slit \Ha\ spectra across N66 showing high- and low-velocity emission corresponding closely to the X-ray boundaries of the supernova remnant.  We use these FUV and optical data to determine the physical parameters of the shock and interaction geometry with N66.  We find that ionizing photons from the many massive cluster stars nearby likely affect the ionization balance in the post-shock gas, hindering the production of lower-ionization and neutral species.  We discuss the importance and inherent difficulty of searching for supernova remnants in or near bright \HII\ regions and suggest that the far ultraviolet provides a viable means to discover and study such remnants.
\end{abstract}


\section{Introduction}
%    background on SNR 0057-7226 and N66
The giant \HII\ region N66 is the largest, brightest star forming region in the Small Magellanic Cloud (SMC).  The populous star cluster NGC\,346 embedded in the nebula is home to several dozen early type stars including a dozen of spectral types O7 and earlier.  Two arcminutes to the east lies the massive, luminous WN+OB system HD\,5980 which underwent a luminous blue variable  (LBV) outburst in 1994 \citep{Koenigsberger95,Barba95}.  These stars are the main source of ionization in the nebula.  

N66 also contains at least one known supernova remnant (SNR), \SNR.  {\it Einstein} X-ray observations showed a bright source in N66 tentatively identified as a SNR \citep{IKT,WangWu92}.  Radio observations by \citet{Mills82} showed bright, extended, thermal emission at 408, 843 and 5000 MHz corresponding to the \HII\ region, but no clear sign of the SNR.  Optical observations also showed only the extremely bright nebular emission but no filamentary structure associated with the SNR.  However, \citet{ChuKennicutt88} discovered faint, high-velocity \Ha\ emission in a long-slit spectrum across N66 and suggested the existence of a SNR.  Subsequently, \citet{YKT} subtracted smoothed \Ha\ data from 843 MHz radio continuum observations and revealed a remnant $\sim$3.2\arcmin\ ($\sim$55 pc at an SMC distance of 59 kpc) in diameter and confirmed the remnant \SNR.  X-ray spectroscopy of \SNR\ with ASCA \citep{Yokogawa00} showed that the spectrum can be modeled as a hot thermal plasma.

High-resolution X-ray observations of N66 with the {\it Chandra X-ray Observatory} \citep{Xmega} show a 130\arcsec$\times$100\arcsec\ region of extended, center-bright, thermal emission corresponding to the SNR.  These high-resolution data show \SNR\ to have relatively uniform surface brightness and no clear temperature gradient from center to rim.  Difference imaging techniques applied to recent radio observations \citep{Filipovic03} at 1.42, 2.37, and 4.80 GHz (20, 13 and 6 cm) show \SNR\ to be a non-thermal, limb-brightened shell with a spectral index of approximately $\alpha$=$-$0.17 (S$_\nu\propto\nu^\alpha$).  The X-ray and radio emitting regions are similar in extent at a resolution of $\sim$20\arcsec.  Hence, the \citet{YKT} estimate of angular size is probably too large, and the X-ray/radio SNR is roughly 36 $\times$ 28 pc in extent. Thus, \SNR\ may be a member of the ``mixed morphology'' class of SNRs \citep{RhoPetre98}, with a radio shell and filled-center X-ray emission, although the presence of HD~5980 along the sight line confuses the situation. \citet{Xmega} resolve X-ray emission from HD~5980 itself for the first time, and note similarities with the galactic object $\eta$ Carinae, which has diffuse surrounding X-ray emission \citep{Sew81}.  

% FUV measurements of SNR0057-7226 -- history continued
\citet{Xmega} mention a possible association of HD~5980 and \SNR, but FUV absorption studies have clearly determined that HD~5980 is behind the SNR.  FUV observations toward HD\,5980 show the expected Galactic and SMC absorption at $\rm v_{lsr}\approx$0 \kms\ and $\rm v_{lsr}\approx$$+$150 \kms, respectively, as well as high-velocity absorption indicating the receding shell of an intervening SNR.  Absorption at $+$300 \kms\ was found in IUE spectra by \citet{deBoerSavage80}; this was later confirmed by \citet{FitzpatrickSavage83} who suggested an intervening SNR.  \citet{Koenigsberger01} detected absorption systems at $+$300 and $+$330 \kms\ in \CIV, \NV\ and several other ions in high-resolution {\it HST} STIS spectra.  They also see very weak components at $+$20 and $+$50 \kms, possibly consistent with the approaching side of the remnant if indeed these features are real.   \citet{Hoopes01} also see absorption at $\sim$$+$300 \kms\ in \CIII\ and \OVI\ in {\it FUSE} observations of HD\,5980 and conclude that the star must lie behind the receding shell of \SNR.  \citet{Hoopes01} also find that a {\it FUSE} spectrum of Sk80 (AV232), an O7Ia star $\sim$1\arcmin\ southeast of HD\,5980 but also seen in projection within the X-ray emission, shows none of the high-velocity absorption seen toward HD\,5980.  Since Sk80 lies closer to the X-ray edge of the remnant, we might expect a lower line-of-sight velocity for SNR gas in this sight line.  However, the \OVI\ absorption profile near the SMC systemic velocity ($\sim$150 \kms) is indistinguishable from that of HD\,5980.  Thus, Sk80 must either lie within or in front of the remnant.

%    importance of OVI
Far-ultraviolet (FUV) observations are useful in the study of SNRs.  Several strong emission lines including \CIV\ $\lambda\lambda$1548,1550, \NV $\lambda\lambda$1238,1242, and especially \OVI\ $\lambda\lambda$1032,1038 are good shock diagnostics and are produced in regions where the temperature is lower than typical X-ray producing regions and higher than regions of bright optical emission \citep{SutherlandDopita93}.  \OVI\ emission, in particular, arises in gas at temperatures near $3\times10^5$ K--a condition almost never reached by photoionization--and is thus a sensitive tracer of shock-heated material.

%    importance of emission/absorption comparison
In this paper we present a {\it FUSE} observation of emission at the X-ray edge of \SNR\  that shows strong emission in \OVI\ $\lambda\lambda$1032,1038 and \CIII\ $\lambda$977.  We also present new longslit echelle observations showing high- and low-velocity material in \Ha\ across the face of \SNR.  The observations are described in \S2.  In \S3, we discuss the SNR kinematics as revealed by the optical and FUV data.  In \S4, we compare the \OVI, \CIII, and \Ha\ measurements with the predictions of shock models to derive shock velocity, preshock density and ram pressure at one location in the SNR.  Using the same models with the FUV absorption data of \citet{Hoopes01} and \citet{Koenigsberger01}, we derive similar values for material on the sight line toward HD\,5980.  We propose a physical picture for \SNR\ in relation to N66.  Finally we discuss the implications these observations have for the detection of SNRs in bright nebulae and OB associations.  A summary is presented in \S5.

%%%%%%%%%%%%%%%%%%%%%%%%%%%% OBSERVATIONS %%%%%%%%%%%%%%%%%%%%%%%%%%%%%

\section{Observations}

%\subsection{Supporting Images} %%%%%%%%%%%%%%%%%%%%
Figure~1 shows images of \SNR\ and its surroundings.  The left panel shows an \Ha+[\NII] emission-line image taken with the Curtis Schmidt Telescope at CTIO in 1999, December.  These data are described more thoroughly by \citet{Danforth02}.  The right panel shows a recent X-ray image of NGC\,346 from the {\it Chandra X-ray Observatory}, obtained courtesy of \citet{Xmega}.  These data show \SNR\ clearly as a bright, extended source.  Unfortunately, the ACIS detector gaps partially obscure the southwestern edge of the SNR and some of the core of NGC\,346 itself.  

\subsection{{\it FUSE} Observations}Six stars within N66/NGC\,346 are part of the {\it FUSE} Hot ISM Team program and are marked with small boxes in Figure~1.  Four of these targets are within the crowded core of the cluster itself.  The cluster targets were observed with the 4\arcsec$\times$20\arcsec\ MDRS aperture.  The remaining two (HD\,5980 and Sk80) lie about two arcminutes to the east of the cluster in projection toward \SNR\ and were observed with the 30\arcsec$\times$30\arcsec\ LWRS aperture.  

{\it FUSE} has three spectral apertures \citep{Moos00}; the LWRS and the MDRS apertures are centered 209\arcsec\ from each other.  Thus, while the MDRS aperture was being used to observe stars in NGC\,346, the LWRS aperture fell serendipitously on other parts of the sky.  In one case (P20305, henceforth Position~1), the LWRS aperture fell 75\arcsec\ east of HD\,5980 on the eastern X-ray edge of \SNR.  This spectrum shows bright emission lines of \CIII\ $\lambda$977.020 and \OVI\ $\lambda\lambda$1031.926, 1036.617.  A very faint continuum underlies the bright emission lines at a flux level of $\sim10^{-14}$ \fluxd.  In another case (P20302, henceforth Position~2) the LWRS aperture fell in a region north of HD\,5980 outside the boundaries of the SNR's X-ray emission.  This sight line shows a weak continuum spectrum similar to that at Position 1.  We use this sight line to characterize the FUV background of N66.  The {\it FUSE} observations are summarized in Table~1.  Figure~1 shows the positions of the {\it FUSE} LWRS apertures for Positions 1 and 2 as large boxes.

The third {\it FUSE} aperture, the 1\farcs25$\times$20\arcsec\ HIRS aperture, lies halfway between the other apertures.  In several observations of stars in NGC\,346 the HIRS aperture fell on X-ray bright regions of \SNR.  However, no flux was detected in these spectra. Since the HIRS aperture has less than 3\% of the collecting area of LWRS for detecting diffuse emission, this negative result is consistent.

% data processing.
{\it FUSE} has four independent spectral channels (in {\it FUSE} nomenclature, LiF1, LiF2, SiC1, and SiC2), and because of thermal effects onboard the spacecraft, relative channel alignments shift in a complicated manner \citep{Moos00,Sahnow00}.  This can cause significant complications for diffuse source observations \citep[cf.][]{Blair02}.  The spacecraft guides on the LiF1 channel, which also has the highest FUV throughput near 1035\AA.  The other three channels tend to drift as much as 8\arcsec\ over the course of a few orbits with respect to LiF1 if no realignments are made.  For this reason, we use only LiF1 data for stellar targets whenever possible.  LiF1 does not cover wavelengths below 987\AA.  Thus we use the lower-throughput SiC2 channel to observe \CIII\ $\lambda$977. All four channels cover the 1035\AA\ region containing \OVI, so a direct comparison of \CIII\ and \OVI\ is possible.  The six stellar datasets were reduced using the methods described by \citet{Danforth02} and references therein.  Spectral resolution in these data is 15--20 \kms.

The LWRS data were reduced using {\sc calfuse} v2.0.5 and observations were combined by channel.  The region around 1032\AA\ was examined in all four channels to determine the relative shifts that exist in the wavelength solutions relative to LiF1.  We increased the signal-to-noise ratio by coadding the two LiF channels to create a composite \OVI\ spectrum.  This does not significantly degrade the spectral resolution, which for an aperture-filling extended source is $\sim$0.34\AA\ (106 \kms\ at 1032\AA).  \OVI\ flux is present in the two SiC channels as well and closely resembles that in the LiF channels, but with lower S/N.  The two SiC channels were coadded for the \CIII\ emission.  Figure~2 shows these data in comparison to Position~2 which sampled a sight line outside the SNR.

\subsection{CTIO 4-m Echelle Spectra} %%%%%%%%%%%%%%%%%%%%
Seven longslit echelle spectra of the N66 region were obtained on the 4-m Blanco telescope at the Cerro Tololo Interamerican Observatory (CTIO) during 2000 December and 2001 December.  These observations are summarized in Table~2 and the slit positions are shown in Figure~1 as dashed lines.  For both observing runs, a 79 line mm$^{-1}$ echelle grating and long-focus red camera were used.  The decker was opened to its maximum extent giving a useful slit length on the sky of 3\farcm5.  The slit width was set to 1\farcs65.  The cross-dispersing grating was replaced with a flat mirror and an order-separating filter ($\lambda_c$=6580\AA, $\Delta\lambda$=150\AA) was used to isolate the spectral region around the \Ha\ line.  [\NII] $\lambda\lambda$6548,6583 also falls within the spectral range sampled.

Data reduction was performed with IRAF\footnote{IRAF is distributed by the National Optical Astronomy Observatories, operated by the Association of Universities for Research in Astronomy, Inc., under cooperative agreement with the National Science Foundation.} using the {\sc quadproc} and {\sc longslit} packages.  Trim, bias and dark current calibration were performed first.  The images were next processed to remove the tilt of the CCD axis with respect to the slit.  This rendered the stellar spectra horizontal along a row.  Thorium-Argon lamp spectral lines were traced and fit to correct for curvature in the cross-dispersion direction and provide a dispersion solution.  Final wavelength calibration was performed using the terrestrial airglow lines present in the observations themselves.  The result is a rectified spectrum with dispersion along the x-axis and the spatial dimension along the y-axis.  The calibrated data cover $\sim$3\farcm5 of sky at a spatial resolution of $\sim$1\arcsec\ (0\farcs267 pix$^{-1}$) and spectral resolution of $\sim$10 \kms\ (0.082\AA\ pix$^{-1}$ = 3.7 \kms\ pix$^{-1}$).  We present three of the spectra in two-dimensional format in Figure~3, where the wavelength scale has been converted to velocity.

%%%%%%%%%%%%%%%%%%%%%%%%%%%%%%% ANALYSIS %%%%%%%%%%%%%%%%%%%%%%%%%%%%%%

\section{Analysis}

\subsection{FUV Emission}
%    Strong OVI and CIII emission from Pos1 ap
The LWRS aperture of Position~1 lies 75\arcsec\ east of HD\,5980 and coincides with the edge of bright, soft X-ray emission (Figure~1).  Strong \OVI\ emission is seen in both lines of the doublet as well as \CIII\ $\lambda$977.02 (Figure~2).  We detect no other lines in the {\it FUSE} wavelength range (905--1187\AA) besides airglow.  In addition to the emission lines, a faint continuum is seen in the Position~1 spectrum (F$_\lambda\sim10^{-14}$~\fluxd).  The spectrum is consistent with an early-type star and shows many of the same ISM absorption features seen in other {\it FUSE} spectra.  The unidentified star in the LWRS aperture visible in the optical image (Figure~1) is not the source of the FUV continuum.  It does not appear in FUV UIT ($\lambda_0=1550$\AA) images of the region \citep{Parker98} and is of spectral type A or later.

In contrast to Position~1, there are no \OVI\ and \CIII\ emissions detected at Position~2.  This aperture fell an arcminute northeast of the edge of the SNR in a field empty of bright stars, UV-bright or otherwise.  However, a similar stellar continuum can be seen in this spectrum.   The faint continuum observed at both positions is due to dust-scattered starlight from the many FUV-bright sources nearby.  A similar effect at longer UV wavelengths was seen in IUE spectra of other regions in the SMC \citep{Blair89}.

% fitting the OVI emission
Any emission from the SNR will suffer absorption from the intervening SMC and Galactic ISM.  Fortunately, there is little molecular hydrogen absorption in the vicinity of 1032\AA; the only ISM absorption comes from SMC and Galactic \OVI.  We use the continuum-normalized \OVI\ absorption profile of Sk80 \citep[their Figure~80]{Danforth02} as our absorption model.  A Gaussian profile divided by the absorption model was convolved with a 106 \kms\ aperture function.  The Gaussian parameters were varied and the resulting models were fit to the observed $\lambda$1032 line profile.  The best fit parameters were: $\rm v_{lsr}$=228 \kms, FWHM=159 \kms, and peak flux=2.23$\times10^{-13}$ \fluxd.  The total flux of the intrinsic Gaussian function is 1.27$\times10^{-13}$ \flux\ compared with an observed flux of 8.68$\times10^{-14}$ \flux.  Both the fit to the data and the intrinsic Gaussian profile are shown in the top panel of Figure~4.  

The \OVI\ $\lambda$1032 line should be twice the intensity of the $\lambda$1038 line in the absence of resonance scattering and self-absorption.  However, this ratio can vary as a function of velocity within the line profile.  The observed $\lambda$1032/$\lambda$1038 ratio is close to two on the blue-shifted side of the line but becomes significantly higher on the red-shifted wing.  Since SMC and Galactic \OVI\ absorption affect both lines in a proportional manner, we conclude that resonance scattering and self-absorption is minimal but that significant absorption from molecular hydrogen---5-0(P1)\HH\ $\lambda$1038.156 and 5-0(R2)\HH\ $\lambda$1038.690---affects the weaker line.  

We can characterize the $\lambda$1038 emission using the intrinsic emission profile determined from the $\lambda$1032 line.  In this case, the absorption model was allowed to vary; a simulated molecular absorption spectrum was calculated with the {\it FUSE} spectral simulator code {\sc fsim} and the Sk80 $\lambda$1032 \OVI\ absorption profile (scaled by a square root) was multiplied in.  (Since $\tau_{1038}=\frac{1}{2}\tau_{1032}$ in the optically thin limit, $\rm I(1038)=I_oexp(-\frac{1}{2}\tau_{1032})=I(1032)^{1/2}$.)  The resulting model was multiplied by the intrinsic Gaussian profile determined above and convolved with the aperture function to simulate the observed $\lambda$1038 line.  

We first fit the $\lambda$1038 emission using two molecular hydrogen components: a Galactic component \citep{Shull00} (v=0 \kms, N(\HH)=$\rm1.6\times10^{16}~cm^{-2}$, b=5 \kms, T=150 K) and an SMC component (v=115 \kms, N(\HH)=$\rm 2\times10^{15}~cm^{-2}$, b=15 \kms, T=400 K) slightly modified from \citet{Tumlinson02}.  However, this did not produce enough absorption on the red side of the emission.  A third component, representing swept-up material from the SNR at v$\sim$155 \kms\ improves the fit (Figure~4, middle panel), but requires an unphysically large turbulent velocity for a good fit.  There could be many components at a range of velocities in the turbulent, post-shock gas which might account for the enhanced absorption.  No such high-velocity components are seen by \citet{Tumlinson02}.

% CIII emission
The \CIII\ $\lambda$977 emission is strongly affected by resonance scattering and self-absorption.  The {\it FUSE} spectrum shows that the Galactic and SMC components are saturated \citep{Danforth02}.  Therefore we do not attempt to fit the observed \CIII\ profile.  The measured flux, 6.3$\times10^{-14}$ \flux, is a lower limit to the intrinsic flux.  Overlaying the derived \OVI\ $\lambda$1032 profile (Figure~4, bottom panel) provides a first order assessment of how this line might be affected if its intrinsic line width is similar to \OVI.

% lack of other emission
Apart from \OVI\ and \CIII\ (and geocoronal airglow), there are no other lines detected.  In a spectrum of the Cygnus Loop, a typical middle-aged SNR, the next brightest lines  to \OVI\ and \CIII\ were [\ion{Ne}{5}] $\lambda$1146 and \ion{S}{4} $\lambda\lambda$1063,1074 at $\sim$5\% of the observed \OVI\ flux \citep{Blair02}.  Such weak emission would not be detected in the spectrum of \SNR, especially considering the lower abundances of these elements in the SMC.  

\subsection{\Ha\ and [\NII] Emission}

% describe the data
The seven echelle observations cut across much of the N66 nebulosity in Figure~1 including the core of NGC\,346 and the outlying regions.  Three spectra cross significant portions of the SNR and show faint high- and low-velocity emission.  The data show both \Ha\ and [\NII] emission at a velocity resolution of $\sim$15 \kms.

%      surprisingly quiescent over main nebula
We find the nebula itself to be surprisingly quiescent on the large scale given the violence of strong stellar winds and ongoing star formation.  The bright nebular emission can be fit with a single-component Gaussian with v$\rm_{lsr}$=157.7 \kms, FWHM=34.7 \kms.  Assuming a temperature of $10^4$ K in the photoionized gas, we derive a bulk turbulence of 23.8 \kms (b=14.3 \kms).  This degree of turbulence is typical of Magellanic \HII\ regions but much lower than that seen in giant \HII\ regions such as 30 Doradus and nebulae in M101 and M33 \citep{ChuKennicutt94,SmithWeedman73,SmithWeedman71,SmithWeedman70}.  The thermal FWHM of [\NII] is narrower than that of \Ha\ by the square root of the ratio of their ionic masses (i.e. by $\sqrt{14}$=3.7).  Thus [\NII] emission is a more sensitive probe of velocity structure than \Ha.   However, these lines are usually much weaker than \Ha, particularly in the low-nitrogen SMC ISM, and are often completely undetectable in SMC echelle spectra.  Four of the echelle spectra (both HD\,5980 pointings, NGC\,346-WB3+4 and NGC\,346-WB6) have usable [\NII] data.  These show $\rm v_{lsr}=158.5$ \kms\ and a $\rm FWHM$=22.8 \kms.  Assuming a temperature of 10$^4$ K, we derive a bulk turbulence of 17.5 \kms\ (b=10.5 \kms), also typical for ordinary Magellanic \HII\ regions. 

%      small expansion feature corresponding to NGC346 rim
The two spectra which cut through NGC\,346 show an expansion feature bounded by the bright \Ha\ rim visible in Figure~1 surrounding the cluster core.  (NGC\,346-WB1 samples the same region as the latter two spectra, but has a high sky brightness masking any [\NII] emission.)  The data show a maximum expansion velocity of 14 \kms.  This expansion is presumably driven by the stellar winds of the stars located within the bubble.  The NGC\,346 expansion feature is also detected in the \Ha\ data, but only in that a pair of Gaussians are needed to fit the broad \Ha\ profile, rather than a single component.  Meanwhile, the east-west HD\,5980 data cut through N66C, a compact \HII\ region to the west of HD\,5980 and north east of the cluster core.  The [\NII] data reveal an expansion feature here with $\rm v_{exp}\approx13$ \kms.  

%      high velocity Ha emission
High- and low-velocity \Ha\ emission appear in the four echelle observations which cross \SNR\ (HD\,5980 NS and EW, Sk80 NS and EW).  Three of these spectra are shown in Figure~3; the fourth spectrum (Sk80 EW) is of low signal-to-noise and intersects only the edge of the remnant.  The faint, higher velocity emission in these spectra arises in shocks associated with the SNR.  It is weaker than the photoionized nebular emission by a factor of $\sim$70 and has a very low signal to noise ratio.  The SNR emission is quite patchy:  blobs with widths of $\sim$10\arcsec\ ($\sim$3 pc) are seen in the position-velocity diagrams (Figure~3).

We find high-velocity features all across the face of \SNR\ with higher velocity emission typically toward the center suggesting an expanding shell.  The fastest emission is at v$\approx+$335 \kms\ about 15\arcsec\ (4 pc) south of HD\,5980 (this is also the region of peak X-ray brightness).  If the systemic velocity of the SNR is taken to be the same as the N66 ISM, this gives an expansion velocity of $\sim175$ \kms, consistent with strong \OVI\ production by the shock front.  

%      low velocity Ha emission
The \Ha\ emission from the approaching side of the SNR is also seen in Figure~3, though it is even fainter than the receding side.  In the few cases where the blue-shifted material is discernible from the Gaussian wings of the bright nebular emission, we see velocities as low as v$\rm_{lsr}\sim+50$ \kms.  As with the receding side of the SNR, the material with the largest velocity offsets tends to be closest to the center of the remnant.

%      flux calibration and surface brightness
Approximate spectrophotometry was performed on the echelle data using a flux-calibrated \Ha+[\NII] image of N66 obtained from \citet{KennicuttHodge86}.  Comparing profiles from the image with wavelength-integrated intensity along the slit gave a conversion factor of $\sim$4.5$\times10^{-18}$ \flux\ count$^{-1}$.  We use this value to measure fluxes and surface brightnesses for different patches of SNR emission in the spectra (Figure~3).  A Gaussian profile was fit to the main nebular emission (v$\approx$158 \kms) over a range $\Delta d$ along the slit and divided from the data.  A background level was then subtracted and the remaining counts within a wavelength (velocity) range $\lambda_1$-$\lambda_2$ were summed.  The summed counts were converted into a total flux.  Surface brightness was determined by dividing the total flux by the area covered by the extraction region ($\rm \Delta d\times slit width$).  Typical observed surface brightnesses were $\sim10^{-16}$ \surfb\ with a standard deviation of 50\%.

%      emission velocities match Ha emission velocity
There is a faint patch of \Ha\ emission at 215 \kms\ at the Position~1 LWRS aperture (Figure~3, top panel).  This corresponds reasonably well with the observed \OVI\ emission at 228 \kms.  We measure the surface brightness of this emission patch to be $8.5\pm2.4\times10^{-17}$ \surfb\ over a slit range 57\arcsec--79\arcsec\ east of HD\,5980.

%%%%%%%%%%%%%%%%%%%%%%%%%%

\section{Discussion}  % S4

The detection of \OVI\ emission as well as the kinematics present in the \Ha\ echelle data imply the presence of a strong shock in N66 associated with \SNR.  The SNR shock model was also favored by \citet{Hoopes01} for the high velocity absorption towards HD\,5980.  Detection of the front side of \SNR\ has not been clearly established because the low velocity components cited by \citet{Koenigsberger01} were detected at such a marginal level. In any event, the near side of the SNR is considerably more difficult to detect both in emission and absorption than the back side.  Optical emission from the \HII\ region complicates the situation and masks the morphology of the SNR. 

In the following discussion we adopt a working model of the physical association of various components in the N66 region.  A roughly spherical SNR is located on the near side of N66.  The rear side of the SNR is propagating into relatively denser nebula while the near side is propagating through the more rarefied SMC ISM.  HD\,5980 lies behind the remnant, embedded within N66.  The NGC\,346 cluster stars lie outside the SNR, also embedded within N66.  We determine the shock velocity, preshock hydrogen density and ram pressure of the shock wave using the emission and absorption sight lines toward \SNR.  By examining the surface brightness ratio between \OVI\ and \Ha, we measure the ``completeness'' of the shock.  Then we discuss the relationship between the SNR and its surroundings and the implications of this analysis for observing SNRs in other \HII\ regions.

\subsection{Physical Properties}  % S4.1

In the N66 reference frame (v$\rm_{lsr}$=$+$158 \kms), the material at Position~1 has a radial velocity of $+$70 \kms\ while toward HD\,5980, $\rm v_r=+142$ \kms.  From X-ray and radio images, we choose the center of the remnant to be (J2000) $\alpha$=00:59:27, $\delta$=$-$72:10:15, about 20\arcsec\ south of HD\,5980.  This position is also consistent with the location of the highest-velocity \Ha\ emission (Figure~3).  Using the projected distances from this center and the two radial velocity measurements, we derive a radius of curvature for the shock front of 80$^{+19}_{-10}$\arcsec\ and an expansion velocity for the \OVI\ gas of v$\rm_{exp}=147\pm7$ \kms.  This places the observed X-ray edge on the eastern side of the shock front inclined $\sim$60\dd\ to our line of sight.  The expansion velocity of 147 \kms\ is not the shock speed, but rather the bulk velocity of the postshock material which should be between 0.75 and 1.0 times the shock speed.  

First we caculate shock models for comparison against the observations.  The models were calculated using an updated version of the code described by \citet{Raymond79}.  The main input parameters are the shock velocity, the preshock density and the elemental abundances.  We ran models for shock velocities between 140 and 200 \kms\ spaced by 10 \kms.  A preshock hydrogen number density of 1 cm$^{-3}$, and SMC elemental abundances \citep{RussellDopita90} were used in all the models.  The calculation is followed until the the recombination zone is complete and the gas temperature has reached about 1000 K.  We present the intensities of selected lines from the shock models in Table~3.  The intensities are relative to \Ha=100 and the flux from nearby multiplet lines of an ion have been summed.  The \OVI\ line flux produced by a shock increases sharply as a function of shock velocity in the range 150 -- 200 \kms.  Shocks below $\sim$160 \kms\ do not produce much \OVI\ and thus this represents a lower limit to the actual shock velocity in the SNR.  Over this same range of shock velocities, the \CIII\ flux remains nearly constant.  The changing ratio of I(\OVI)/I(\CIII) is due to the increasing ionization of oxygen at higher velocities. 

As discussed in \S3.1, the \CIII\ is strongly affected by absorption and scattering, and so it is not possible for us to derive an accurate assessment of the intrinsic ratio of I(\OVI)/I(\CIII) in the spectrum of Position~1.  However, we note that using the total (corrected) \OVI\ flux with the observed \CIII\ flux places an upper limit of F(\OVI)/F(\CIII)$\le$3.  Ignoring the effect of differential reddening for the moment, this clearly points us toward the lower end of the velocity range where \OVI\ is produced.  Based on this ratio, the observed expansion velocity and the shock velocity requirements for \OVI\ production, we adopt 160 \kms\ as a representative shock velocity at Position~1 and use details from the model at this velocity below.

We now compare the observed and calculated \OVI\ surface brightnesses to estimate the preshock density required to produce the observed \OVI\ flux.  The model predicts an \OVI\ surface brightness (in both lines of the doublet) of I$\rm_{mod}=6.08\times10^{-5}$ \flux\ into 2$\pi$ steradians (for v$\rm_{shock}$=160 \kms, $\rm n_o$=1 cm$^{-3}$, He/H=0.08) or I$\rm_{mod}=2.28\times10^{-16}$ \surfb.  In the optically thin limit, intensity will scale with density so we can write

\[ \frac{I_{obs}~C_{1035}}{A}~=I_{mod}~n_o\]

\noindent where $C_{1035}$ is the reddening correction factor at 1035\AA, $A$ is the surface area of shock (in arcseconds$^2$) within the aperture and $n_o$ is the preshock density in units of cm$^{-3}$.  I$\rm_{obs}$ is the total \OVI\ flux in both lines which we estimate by multiplying the corrected \OVI\ $\lambda$1032 flux by 1.5, thus I$\rm_{obs}=1.91\times10^{-13}$ \flux.  From Figure~1, we estimate that $\sim$$\frac{1}{3}$ of the LWRS aperture is filled with X-ray emission.  If the \OVI\ emission follows the same pattern and the shock front is tilted 60\dd\ from our line of sight, then the aperture sampled emission from 600 arcseconds$^2$ of shock front.

Reddening is a potential source of uncertainty.  \citet{MPG} found the mean E(B$-$V)=0.14 for the NGC\,346 stellar population.  However, since the SNR lies in front of the bulk of the cluster, this may overestimate the extinction to the SNR.  On the other hand, using the spectral type and observed colors of Sk80, one can derive E(B$-$V)=0.11.  We adopt this value and assume R$\rm_{v}$=3.1 (typical for the diffuse ISM) and the extinction curve of \citet{HutchingsGiasson01} (E(1035\AA$-$V)/E(B$-$V)=18).  This yields a reddening correction C$_{1035}$=4.5.  

Solving the above equation, we get n$\rm_o=6.3$ cm$^{-3}$ at Position~1, similar to an azimuthally-averaged electron density of 7 cm$^{-3}$ at the radius of Position~1 derived by \citet{Kennicutt84}.  The dynamical pressure is then $\rho~v_s^2=1.3~m_H~n_o~v_s^2=3.5\times10^{-9}$ dyne cm$^{-2}$.  If more reddening is assumed and E(B$-$V) is raised to 0.14 (C$_{1035}$=6.8), the changes are minor:  n$\rm_o=9.5$ cm$^{-3}$ and $\rm P=5.2\times10^{-9}$ dyne cm$^{-2}$.  Less reddening (E(B$-$V)=0.08, C$_{1035}$=3.0) would yield n$\rm_o=4.2$ cm$^{-3}$, $\rm P=2.3\times10^{-9}$ dyne cm$^{-2}$.  

This pressure is higher than those seen in regions of the Vela SNR.  \citet{Sankrit01} derive n$\rm_o=0.5$ cm$^{-3}$ and P=3.7$\times10^{-10}$ dyne cm$^{-2}$ for a X-ray bright, nearly face-on shock.  \citet{Raymond97} find P=1.6$\times10^{-9}$ dyne cm$^{-2}$ for an edge-on shock.  \citet{JenkinsWallerstein95} find P=2-4$\times10^{-10}$ dyne cm$^{-2}$ for another locations in Vela while \citet{Kahn85} find P$\sim$1$\times10^{-10}$ dyne cm$^{-2}$ for a region of diffuse X-ray emission.  This variation in pressures within the same remnant suggests local density and pressure enhancements caused by reverse shocks from blastwave-cloud interactions.  We have only two widely-separated measurements for \SNR\ and can say nothing so specific but the high pressure and density at Position~1 as well as the X-ray emission morphology suggest similar shock-cloud interactions.

% conditions toward HD5980
The shock models discussed above predict the total column depths of ions produced in the post-shock flow.  \citet{Hoopes01} measured the column depths of several highly ionized species in the high-velocity component toward HD\,5980 (their Table~1) and conclude that a SNR shock is the most likely production mechanism for these ions in the observed quantities.  \citet{Koenigsberger01} presented data on other ions not available to {\it FUSE}.  In Figure~5, we present the observed column depths of these FUV ions and the values predicted by the models with v$\rm_{s}$=160, 180 and 200 \kms.  The ions are ordered according to the ionization potential needed to produce them.  The relative column depths of all the ions down to \FeIII\ are well matched by the 160 \kms\ model.  The higher velocity models predict too much \OVI\ relative to the other high ionization ions.  The column depths scale linearly with preshock density (all the models were run with n$_0$ = 1 cm$^{-3}$).  The offset between the 160 \kms\ model and the observations is $0.78\pm0.14$ dex, so a preshock density of $\rm n_o=6\pm2 cm^{-3}$ can reproduce the observed column depths for the high ionization lines.  The shock velocity and preshock density toward HD\,5980 are thus in good agreement with our measurements for Position~1.

The behavior of the low ionization lines is drastically different.  For \SiII\ and \FeII, all the models predict column densities that are an order of magnitude higher than the observations (see Figure~5).  This dearth of low-ionization material can also be seen by comparing the observed and calculated ratio of \OVI\ and H$\alpha$.  The shock models (which have been calculated to the point of complete recombination) predict I(\OVI)/I(\Ha)=3.3 for $\rm v_s$=160 \kms\ and SMC abundances.  The observed ratio at Position~1 is 

\[  \frac{I(OVI)}{I(H\alpha)} = \frac{SB(OVI)}{SB(H\alpha)}~\frac{C_{1035}}{C_{6563}}~\frac{f_{H\alpha}}{f_{OVI}} \]

\noindent where $SB$ indicates the surface brightness and $f$ the aperture filling factor of the relevant emission.  We assume the reddening correction $C_{1035}$=4.5 as discussed above; reddening at \Ha\ is moderate and we adopt $C_{6563}$=1.24.  The filling factor $f$ for each aperture is more uncertain.  We assume $f_{OVI}\approx\frac{1}{3}$ as discussed above.  It is difficult to say where emission does and does not appear at a fine scale in the low S/N echelle data.  The \Ha\ aperture was chosen carefully to encompass only the faint emission in Figure~3, but we estimate that $f_{H\alpha}$ is still $\sim$$\frac{2}{3}$.  These values give I(\OVI)/I(\Ha)$\approx$18.  Uncertainties in the filling factors, reddening, and surface brightnesses probably make this ratio uncertain by a factor of 2, but the ratio is still significantly higher than the predicted ratio of 3.3.  

The weakness of \Ha\ emission compared with \OVI\ at Position~1 suggests the presence of an incomplete shock.  There are many cases where incomplete shocks are observed in individual filaments or small portions of SNRs \citep{Fesen82,Blair91,Danforth01}; however, these regions are all at sub-parsec scales.  The LWRS aperture at Position~1 is sampling a region of \SNR\ roughly 8 pc across.  Furthermore, the observed \Ha\ surface brightness does not vary strongly between different portions of the SNR implying that the conditions at Position~1 are typical of the remnant as a whole.  In Figure~3, we see weak SNR \Ha\ emission or none at all.  This is a large object with moderate shock velocities implying that it is a middle-aged remnant like the Cygnus Loop or the Vela SNR; most of the optical emission from these remnants comes from complete, radiative shocks and these remnants are bright in \Ha.  If some portions of \SNR\ were recombinationally complete, we would expect to see \Ha\ emission $\sim$10 times brighter than is observed.  Thus, it appears that there is a global effect preventing the cooler parts of the recombination flow from forming. 

% Photoionization from NGC346 affects the ionization balance in the post-shock material suppressing Ha emission %%%
We argue that the gas at the back end of the post-shock flow is being prevented from recombining by the ionizing flux from the bright stars in N66. There are eleven stars of spectral type O6.5 or earlier located 2\arcmin\ (35 pc, in projection) to the west of \SNR.  The combined ionizing flux from NGC\,346 is 4.0$\times$10$^{40}$ erg s$^{-1}$ \citep{Relano02} or $\sim$1.5$\times$10$^{51}$ ionizing photons s$^{-1}$.  HD\,5980 itself contributes $\sim4\times10^{50}$ ionizing photons s$^{-1}$. Using basic ionization/recombination equations from \citet{Osterbrock89} we find that this photon flux in a medium with density $\sim$5 cm$^{-3}$ will yield an ionized sphere over 100 pc in radius.  Indeed, \citet{Relano02} find that N66 is density bounded and that 45\% of the ionizing photons from cluster stars escape to ionize the general SMC ISM.  This zone of ionization could encompass \SNR\ and affect the ionization balance in the recombining post-shock gas.  The dynamical time scale of the shock is of order a few hundred years.  The photoionization time scales are much shorter ($\sim$100 days), while the recombination times are longer (of order 10,000 yr).  Thus the ionizing flux will suppress \Ha\ emission.  

The strength and spectral energy distribution of the local ionizing flux will affect the column densities of the low and moderately ionized species.  In theory, much of the discrepancy between the modeled and observed \FeII\ and \SiII\ columns towards HD\,5980 (Figure~5) could be due to iron and silicon being ionized to higher ionization stages.  If the best fit model is scaled up to n$\rm_o=6$ cm$^{-3}$, the \FeIII\ column is underpredicted by about 0.3 dex.  This is clearly not enough to explain the 1.9 dex discrepancy in the \FeII\ column, but the stellar spectrum extends out to $\sim$50 eV \citep{Relano02} and the local flux could be producing \ion{Fe}{4}.  However, no such discrepancies are seen in the \SiIII\ and \SiIV\ columns.

A more likely scenario is that the sight line to HD\,5980 is passing through a small region where the swept up column is lower than average for \SNR.  While the \FeIII\ zone is complete, the \FeII\ zone would be incomplete regardless of ionizing radiation.  A self-consistent shock model calculation that includes a photoionizing field is necessary to predict the column densities and the \Ha\ flux accurately.  While such modeling is beyond the scope of this paper, we conclude that the local ionizing field must have a significant effect on the observed characteristics of \SNR\ as a whole.

\subsection{The Relationship Between \SNR\ and N66}  % S4.2

The emission and absorption data discussed here lead to a schematic structure for the N66--SNR region shown in Figure~6.  \SNR\ lies on the near side of N66 and is encountering the denser nebular material on its back side.  Position~1 is located at the edge of the X-ray emission (Figure~1).  The observed radial velocity of \OVI\ and \Ha\ emission at Position~1 is $\sim$225 \kms.  This is lower than the observed velocity of absorption ($+$300 \kms) toward HD\,5980 which lies projected closer to the center of the remnant.  Also it is $\sim$75 \kms\ higher than the systemic velocity of N66 ($+$158 \kms), which we would expect for material viewed tangentially at the edge of the expanding remnant.  Either the systemic velocity of the SNR and N66 differ by $\sim$70 \kms, or the observed X-ray edge (a portion of which is Position~1) does not represent the actual limb of the SNR shock.  The latter is assumed in Figure~6.

We propose that the X-ray emission arises over a spherical cap where the shock has encountered and heated denser material.  The remnant is not yet even half-way submerged in the N66 material, and thus the observed radius of X-ray emission is smaller than the actual blast wave radius.  The observed X-ray edge at Position~1 is inclined to our line of sight by $\sim$60\dd.  HD\,5980 and the other UV-bright stars are located within N66 and have ionized a region of surrounding material.  The combined stellar winds and thermal pressure have formed a slowly expanding envelope around the cluster core.  Since high-velocity absorption is not seen in the spectrum of Sk80 \citep{Hoopes01}, it must lie within or in front of the SNR.  HD\,5980 lies behind the remnant and affects the ionization balance in the swept-up gas, preventing the formation of \HI, \SII, \FeII, \SiII\ and other low-excitation species.  The bright nebular emission seen surrounding the UV bright stars of N66 arises in dense material behind an ionized layer. 

% nearside material 
The approaching SNR material is apparent in \Ha\ in Figure~3, but has not been conclusively identified in FUV spectra.  If the far side of the remnant is seen near $\sim$300 \kms\ and if the SNR is centered at the SMC systemic velocity of $\sim$155 \kms, we expect the nearside absorption to be seen near v=10 \kms.  This region is essentially masked by Galactic absorption making it difficult to detect the SNR.  \citet{Hoopes01} compared the \OVI\ spectra of Sk80 and HD\,5980 near zero velocity and found no difference.  However, both of these sight lines are toward the remnant and may contain nearly identical foreground \OVI\ columns.

The four {\it FUSE} targets in NGC\,346 lie less than 2\arcmin\ from HD\,5980 and are outside the SNR sight line.  A comparison of the averaged, normalized \OVI\ profiles from these four targets and the two sight lines projected within the SNR is shown in Figure~7.  We find a small but systematic excess in \OVI\ absorption near v=0 \kms\ in the two remnant profiles.  The difference represents an extra \OVI\ column density of about $10^{13.8}$ cm$^{-2}$ or about $\frac{1}{4}$ that of the backside value reported by \citet{Hoopes01}.  We also find that the velocity centroid of the HD\,5980/Sk80 zero velocity absorption is shifted by about $+$10 \kms\ with respect to the NGC\,346 Galactic absorption.  This is not too different from the velocity found by \citet{Koenigsberger01} for \SiIV\ and \CIV\ absorption features that they attribute to front-side SNR material.

Recent work by \citet{Howk02} on Galactic halo \OVI\ absorption has shown significant variations in \OVI\ column over as little as 3\arcmin.  The variation in N(\OVI) between the four NGC\,346 targets (mean separation 19\arcsec) is 9\% (the maximum variation between two measurements is 12\%) while the variation between the two SNR sight lines (separation $<$1\arcmin) is only 8\%, comparable to the uncertainties due to continuum placement, noise, and limits of integration.  However, the cluster and SNR stars show a relative difference of 53\% in the zero-velocity \OVI\ column over $\sim$2\arcmin.  Because of the potential Galactic halo variation on small scales, the matter of frontside absorption remains inconclusive.  However, it is plausible we are seeing front side absorption near v=$+$10 \kms\ in the HD\,5980/Sk80 data.  If true, then Sk80 would be inside \SNR, as shown in Figure~6.  Even if it is not, we can place an upper limit of N(\OVI)$\la6\times10^{13}$ cm$^{-2}$ for the frontside \OVI\ column density, much less than is observed from the back side of \SNR.

% response to referee major comment (re: astro-ph/0211491)
In a recent paper, \citet{Velazquez03} model the collision of a SNR shock with a wind-blown bubble around an evolved star and suggest that the far side of \SNR\ is interacting with the stellar wind of HD\,5980.  The X-ray image of the system \citep{Xmega} does, in fact, show some resemblance to the models and there is reasonable agreement between predicted and observed X-ray fluxes.  Our observations allow us to further test this wind-SNR model.  The Vel\'azquez et~al. wind-blown bubble has a radius of $\sim$10 pc while our Position~1 (at the rim of the SNR) is at least twice that distance (in projection) from HD\,5980; thus, Position~1 is sampling the ISM, not stellar wind.  In \S4.1 we derived a preshock density of $\sim$6 cm$^{-3}$ for the material at both Position~1 and in the shock toward HD\,5980, at least two orders of magnitude higher than the interior of the modeled wind bubbles \citep[see Figures~1 and 2 of][]{Velazquez03}.  Furthermore, the $+$300 \kms\ component exhibits an ionization structure expected for a single SNR shock \citep{Hoopes01}.  If this component arose in a turbulent region between SNR and stellar wind shock fronts, as suggested by \citet{Velazquez03}, we might not expect to see the kinematics structure observed by \citet{Hoopes01}.  Clearly, higher resolution spectral observations and detailed shock models are required.

\subsection{Implications for SNRs in \HII\ Regions}  % S4.3

Core-collapse supernovae arise from massive stars often formed in clusters and associations--the same massive stars which photoionize their surrounding gas and create bright \HII\ regions.  Massive SN progenitors cannot migrate far from their birth places in their short lifetimes and thus SNRs associated with H~II regions should occur frequently.  To various degrees, this is not seen in optical SNR surveys of nearby galaxies.

The primary optical method for detecting SNRs, especially in galaxies beyond the Magellanic Clouds, is to compare \Ha\ to [\SII] emission, and pick out objects with [\SII]/\Ha $\ge$ 0.4 \citep[cf.][and references therein]{Long90,BL97}.  In recent years, as the sensitivity and resolution of radio and X-ray observations has become competitive with optical ground-based observations, independent searches at these wavelengths have been undertaken and compared against the optical SNR lists \citep{Lacey97,Gordon98,Gordon99,Pan00,Pan02}.  These multiwavelength surveys and comparisons provide a more complete survey method, but in general do not find many of the same SNRs.  Discussions of selection effects usually point toward ``confusion'' as being responsible for incompleteness of optical SNR surveys in crowded and complex regions of emission.  Clearly this is plausible at some level, but the denser conditions in \HII\ regions might also be expected to cause brighter optical (radiative) SNRs since the shock emissivity increases as the density squared.  Scanning through the optical SNR catalogs of nearby galaxies \citep[to name a few]{BL97,Long90,MF97}, one can find a number of stunning counter examples to the ``confusion'' hypothesis.  There are often bright optical SNRs found in confused regions and adjacent to bright \HII\ region emission.  The mechanism we propose in \S4.2 seems to be a more plausible explanation for why the optical SNR searches miss some SNRs and not others in confused regions.  If a supernova occurs close enough to ionizing stars, the potential bright optical SNR emission in \Ha\ and \SII\ is prevented from forming and no optical SNR can be detected, especially against bright nebular emission.

Our investigation of \SNR, however, demonstrates that bright FUV emission lines are still present in such a situation and may provide an alternate way of not only detecting but studying such SNRs in detail.  Detecting SNRs with higher ions circumvents the problems of truncated recombination and bright background emission.  Stellar photoionization will not prevent the production of high ions nor will they be contaminated significantly by surrounding nebular emission.  Our observations have shown that in a bright \HII\ region, the detection of localized emission from high-ionization species such as \OVI\ can provide a clear signature of SNR emission.  Position~1 lies on the edge of the SNR and shows strong emission in \OVI\ and \CIII.  Position~2 lies outside the remnant but still within bright optical emission; yet no line emission is present.  In conjunction with radio and X-ray data, such a detection is conclusive.  We note however that the use of the FUV is restricted to lines of sight where extinction is low.  Nearby extragalactic targets such as the Magellanic Clouds and Local Group galaxies will be the easiest to investigate with this technique.

%%%%%%%%%%%%%%%%%%%%%%%%%%%%%%%%%%%%%%%%%%%%%%%%%%%%%%%%%%%%%%%%%%%%%%%%%

\section{Summary}   % S5

% FUV emission in Pos1
We see strong emission in \OVI\ and \CIII\ at a position on the eastern X-ray edge of \SNR.  Absorption in these same ions is seen at $\sim$300 \kms\ toward the bright LBV system HD\,5980 which lies behind the remnant.  Another position away from the remnant but within the bright nebular extent of N66 shows no FUV line emission.  The remnant lies on the near side of N66 and the \OVI\ is formed in shocks as the blast wave encounters the denser material of the nebula.

% dynamics of N66
Seven longslit echelle spectra in \Ha\ and [\NII] show that N66 has turbulence similar to other Magellanic \HII\ regions.  Two slowly expanding features are seen: the compact \HII\ region N66C and a bright shell around the core of NGC\,346.  Three of the echelle spectra show faint patches of \Ha\ emission at high and low velocities which appear as a rough expansion feature corresponding spatially and kinematically to \SNR.

Observed \CIII\ and \OVI\ flux compared with shock models allows us to estimate the shock velocity and determine the ambient density and ram pressure at Position~1.  We use observed column depths from \citet{Hoopes01} and \citet{Koenigsberger01} to derive the shock velocity and density of the SNR shock toward HD\,5980.  These values are in good agreement with each other and with other values in the literature.

Weak \Ha\ emission as well as low column densities of low-ionization species such as \FeII\ show that the ionization balance in the post-shock gas is affected by ionizing stellar photons.  The ionizing flux from the many hot stars in NGC\,346 is sufficient to keep swept-up material out of ionization equilibriuim.  This highlights the importance and inherent difficulty of looking for SNRs in bright \HII\ regions.  Direct optical identification is complicated by the bright nebular emission.  FUV emission and absorption, along with X-ray and radio observations, are good ways to circumvent these difficulties.

% Acknowledgments
The authors wish to acknowledge useful discussion with Charles Hoopes, Miroslav Filipovi\'c, Lister Staveley-Smith and Alex Fullerton.  Mike Corcoran provided the X-ray images.  Chris Smith supplied the interference filters used in the Schmidt imagery.  Bryan Dunne helped with the echelle observations.  Robert Kennicutt supplied the flux-calibrated image of N66.  John Raymond made available the shock code used in calculating the models.  We would also like to recognize CTIO for providing the excellent observing facilities and travel support.  This work contains data obtained for the Guaranteed Time Team by the NASA-CNES-CSA {\it FUSE} mission operated by the Johns Hopkins University.  Financial support has been provided by NASA contract NAS5-32985.

%  References

%%% Table 1: FUSE Observations Log %%%

\begin{deluxetable}{llllllcrl}
\tabletypesize{\footnotesize}
\tablecolumns{9}
\tablecaption{{\it FUSE} Observation Summary}
\tablehead{
\colhead{Sight Line}&
\colhead{RA (J2000)}  &
\colhead{Dec (J2000)} &
\colhead{SpType} &
\colhead{FUSEID} &
\colhead{Aper\tablenotemark{a}}  &
\colhead{\# Exp} &
\colhead{ExpTime} &
\colhead{Obs Date}\\
\colhead{}                                          &        
\colhead{{h}\phn{m}\phn{s}}                         &
\colhead{\phn{\arcdeg}~\phn{\arcmin}~\phn{\arcsec}} &
\colhead{}                                          &          
\colhead{}                                          &       
\colhead{}                                          & 
\colhead{}                                          & 
\colhead{(sec)}                                     & 
\colhead{}                                          }
\startdata
NGC\,346-WB6& 00 58 57.74 & $-$72 10 33.6 & O4V((f))   & P20305\tablenotemark{b}&MDRS & 5&10992&2001-Sep-25\\
NGC\,346-WB4& 00 59 00.39 & $-$72 10 37.9 & O5-6V      & P20304\tablenotemark{b}&MDRS & 6&11853&2001-Aug-25\\
NGC\,346-WB3& 00 59 01.09 & $-$72 10 28.2 & O2III(f*)  & P20303\tablenotemark{b}&MDRS & 3& 8482&2001-Aug-25\\
NGC\,346-WB1& 00 59 04.81 & $-$72 10 24.8 & O4III(n)(f)& P20302\tablenotemark{b}&MDRS & 3& 4625&2001-Aug-25\\
HD\,5980    & 00 59 26.55 & $-$72 09 53.8 & WN var  & P10301 & LWRS & 4& 5734                  &2000-Jul-02\\
Sk\,80      & 00 59 31.95 & $-$72 10 45.8 & O7 Iaf+ & P10302 & LWRS & 4&11699                  &2000-Jul-02\\
N66-Pos\,1  & 00 59 42    & $-$72 09 49   & \nodata & P20305 & LWRS & 5&10992                  &2001-Sep-25\\
N66-Pos\,2  & 00 59 37    & $-$72 07 57   & \nodata & P20302 & LWRS & 3& 4625                  &2001-Aug-25\\
\enddata
\tablenotetext{a}{{\it FUSE} Apertures: LWRS=30\arcsec\ square, MDRS=4\arcsec$\times$20\arcsec.}
\tablenotetext{b}{Combined into composite NGC\,346 spectrum.}
\end{deluxetable}

%%% Table 2: Echelle Observation Log %%%
\begin{deluxetable}{lllrrc}
\tabletypesize{\footnotesize}
\tablecolumns{6} 
\tablewidth{0pt} 
\tablecaption{Summary of CTIO 4-m Observations}
\tablehead{
\colhead{Target}                                    &           
\colhead{RA\,(J2000)}                               &           
\colhead{Dec\,(J2000)}                              &            
\colhead{PA}                                        &           
\colhead{ExpTime}                                   &           
\colhead{Date}                                      \\           
\colhead{}                                          &
\colhead{{h}\phn{m}\phn{s}}                         &
\colhead{\phn{\arcdeg}~\phn{\arcmin}~\phn{\arcsec}} &           
\colhead{\dd}                                       &            
\colhead{sec}                                       &            
\colhead{{y}\phn{m}\phn{d}}                          }
\startdata
HD\,5980 NS    & 00 59 26.55 &-72 09 53.8 &   0 & 1200 & 2001-Dec-24 \\
HD\,5980 EW    & 00 59 26.55 &-72 09 53.8 &  90 & 1200 & 2001-Dec-25 \\
Sk80 NS        & 00 59 31.95 &-72 10 45.8 &   0 & 1200 & 2001-Dec-26 \\
Sk80 EW        & 00 59 31.95 &-72 10 45.8 &  90 &  600 & 2000-Dec-10 \\
NGC\,346-WB6   & 00 58 57.74 &-72 10 33.6 &  90 &  600 & 2000-Dec-10 \\
NGC\,346-WB3+4 & 00 59 01.09 &-72 10 28.2 & 197 &  600 & 2000-Dec-10 \\
NGC\,346-WB1   & 00 59 04.81 &-72 10 24.8 &   0 &  600 & 2001-Dec-26 \\
\enddata
\end{deluxetable}

%%% Table 3: Shock Models %%%
\begin{deluxetable}{lrrrrrrr}
\tabletypesize{\footnotesize}
\tablecolumns{8}
\tablewidth{0pt}
\tablecaption{Selected Lines from the Shock Models}
\tablehead{
\colhead{Line / v$_{s}$ (\kms)} &
\colhead{140}            &
\colhead{150}            &
\colhead{160}            &
\colhead{170}            &
\colhead{180}            &
\colhead{190}            &
\colhead{200} 
}
\startdata
\Ha\ ~   6563 &  100 &  100 &  100 &  100 &  100 &  100 &  100 \\
S VI ~     937 &   12 &   19 &   18 &   20 &   23 &   27 &   30 \\
C III ~    977 &  352 &  304 &  275 &  258 &  247 &  239 &  232 \\
N III ~    991 &   17 &   13 &   12 &   11 &   10 &   10 &   10 \\
Ne VI ~   1006 &    0 &    1 &    4 &   12 &   27 &   45 &   58 \\
O VI ~    1034 &    9 &   90 &  334 &  708 &  996 & 1156 & 1236 \\
S IV  ~   1070 &   11 &    8 &    8 &    7 &    7 &    7 &    7 \\
Ne V  ~   1146 &    3 &    8 &   13 &   18 &   24 &   27 &   28 \\
N V  ~    1240 &   17 &   27 &   23 &   23 &   24 &   24 &   24 \\
Si IV ~   1397 &   31 &   28 &   26 &   25 &   24 &   24 &   23 \\
O IV] ~   1402 &  100 &  112 &  108 &  101 &   97 &   94 &   92 \\
N IV] ~   1490 &    9 &    7 &    6 &    5 &    5 &    5 &    5 \\
C IV ~    1549 &  211 &  174 &  156 &  148 &  145 &  142 &  140 \\
O III] ~  1662 &   59 &   52 &   47 &   44 &   42 &   41 &   40 \\
N III] ~  1748 &    6 &    5 &    4 &    4 &    4 &    4 &    4 \\
I(\Ha)\tablenotemark{a}& 1.268 & 1.551 & 1.819 & 2.048 & 2.249 & 2.431 & 2.622 \\
\enddata
\tablenotetext{a}{Flux units are 10$^{-5}$ erg s$^{-1}$ cm$^{-2}$ into 2$\pi$ steradian for n$\rm_o$=1 cm$^{-3}$ and SMC abundances.}
\end{deluxetable}

%%% Figures 1a,b %%%  Optical and X-ray finder charts
\begin{figure}
\plottwo{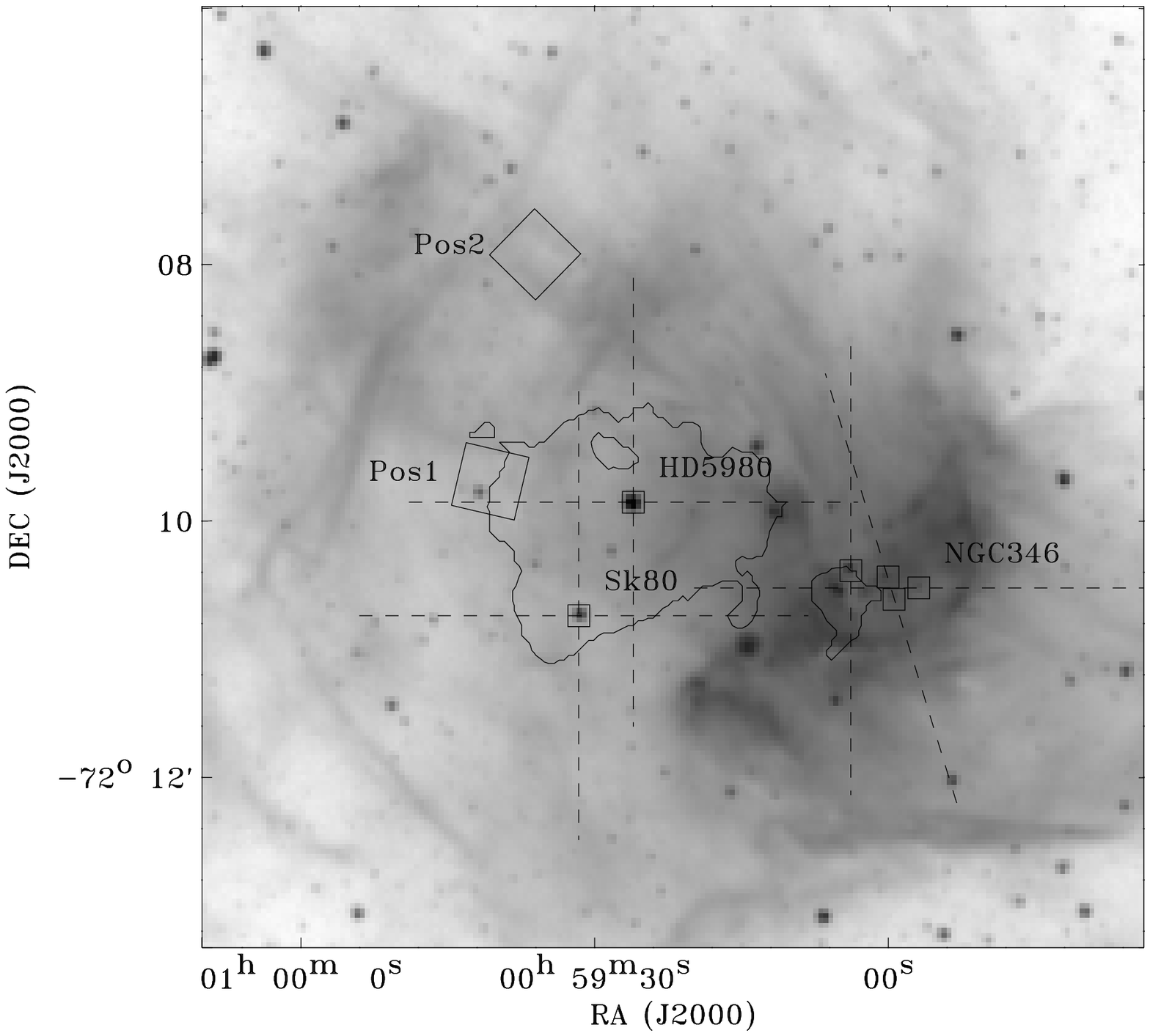}{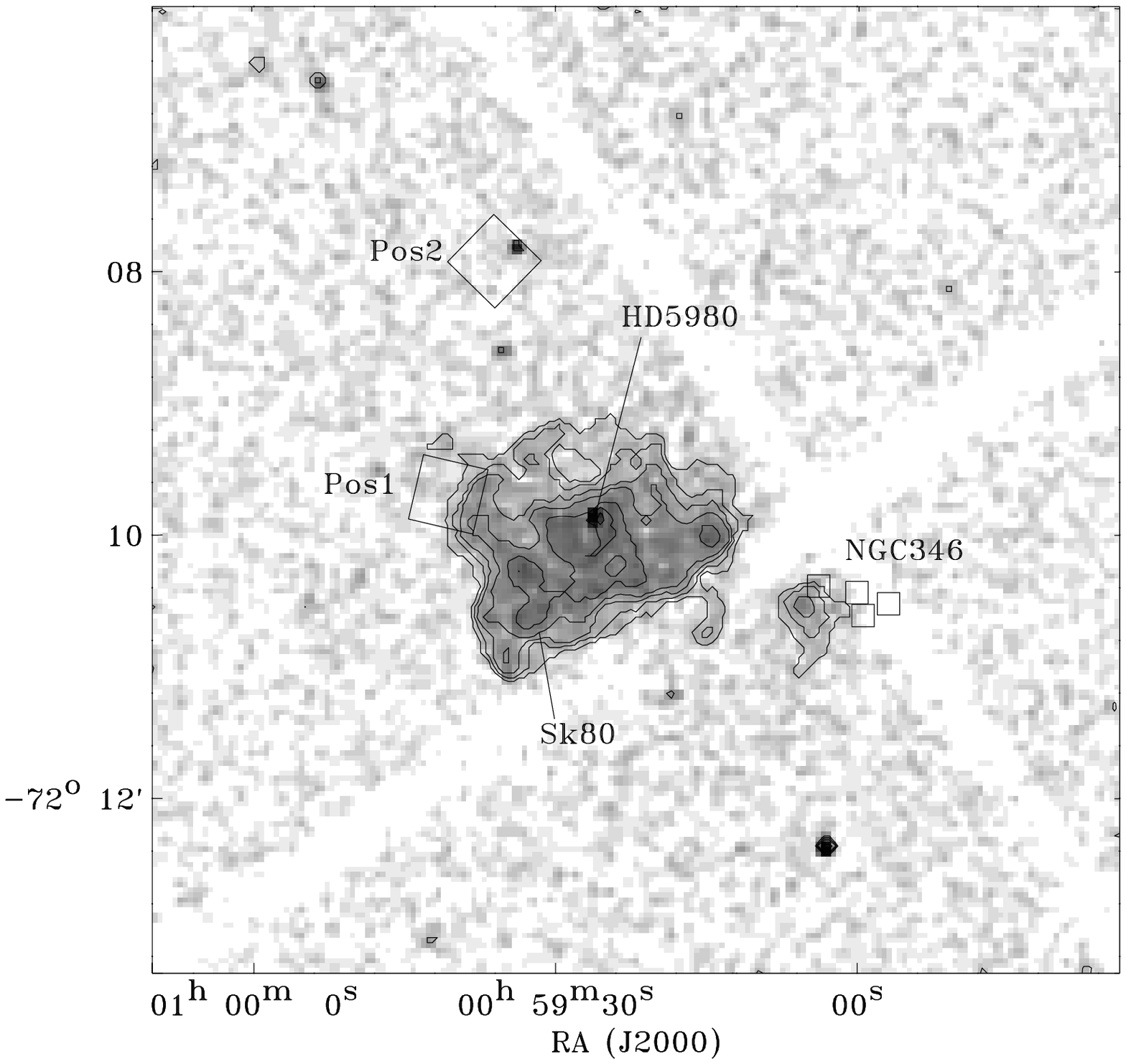}
\caption[N66 in \Ha\ and X-ray wavelengths]
{The giant \HII\ region N66.  An \Ha\ image is shown in the left panel and {\it FUSE} stellar targets are marked with small squares.  LWRS emission apertures are the large boxes in both panels.  Dashed lines show the locations of the echelle spectral slits.  The lowest X-ray contour is also shown.  The right panel shows the {\it Chandra} 0.25-10 keV X-rays \citep[courtesy of][]{Xmega}.  The ACIS chip boundaries can be seen partially obscuring the southern edge of the remnant.}
\end{figure}

%%% Figure 2 %%%  P20305 vs P20302 emission 
\begin{figure}
\epsscale{.75}\plotone{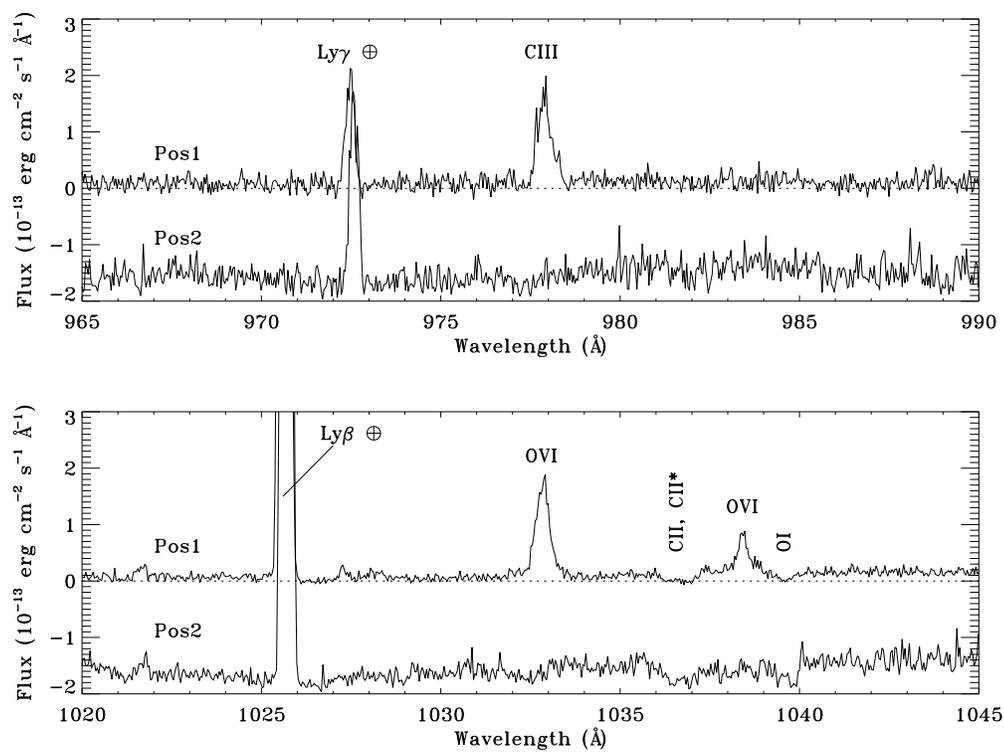}
\caption[\OVI\ and \CIII\ emission from \SNR]
{{\it FUSE} emission spectra at Positions~1 (\SNR) and~2 (N66 background) around 977 and 1032\AA.  Position 2 data has been offset downward by two units.  Note the strong \OVI\ and \CIII\ emission from Position~1 and the lack of emission from Position~2.}
\end{figure}

%%% Figure 3 %%% Echelle data
\begin{figure}
\epsscale{.75}\plotone{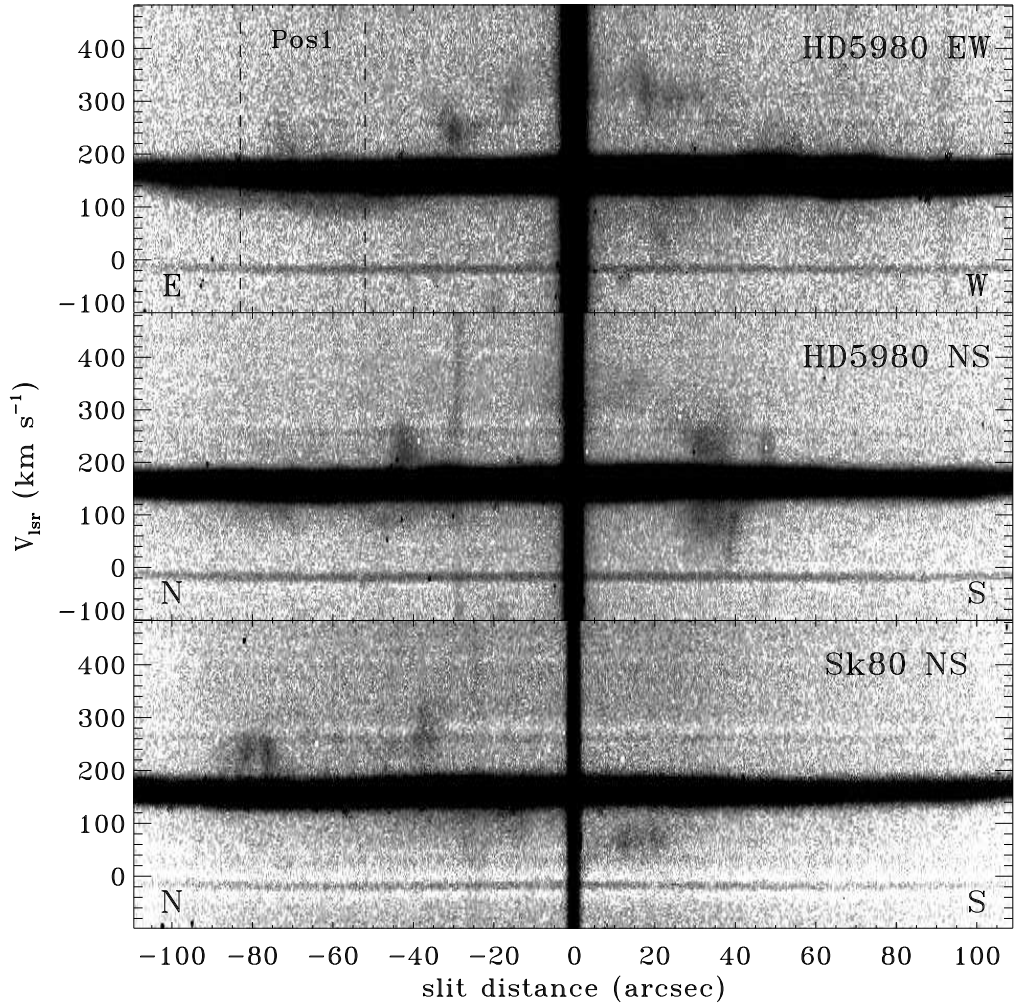}
\caption[Echelle observations of \SNR]
{\Ha\ 6562.82\AA\ echelle observations of \SNR.  The top two slits were centered on HD\,5980 oriented east-west (top) and north-south (middle).  The bottom slit shows a north-south slit centered on Sk80.  Bright nebular emission is seen at v$\sim+$160 \kms\ and the stellar continuum from the indicated star is a vertical band in the center.  Faint, patchy emission at high and low velocities can be seen above and below the systemic velocity along all three slits; this emission arises in the SNR shock.  Terrestrial \Ha\ airglow can be seen as a faint horizontal stripe at v$\sim-20$ \kms.  The location of the Position~1 LWRS slit is shown by the dashed lines in the top panel.}
\end{figure}

%%% Figure 4 %%% P20305 OVI 1032 and 1038 data fits
\begin{figure}
\epsscale{.5}\plotone{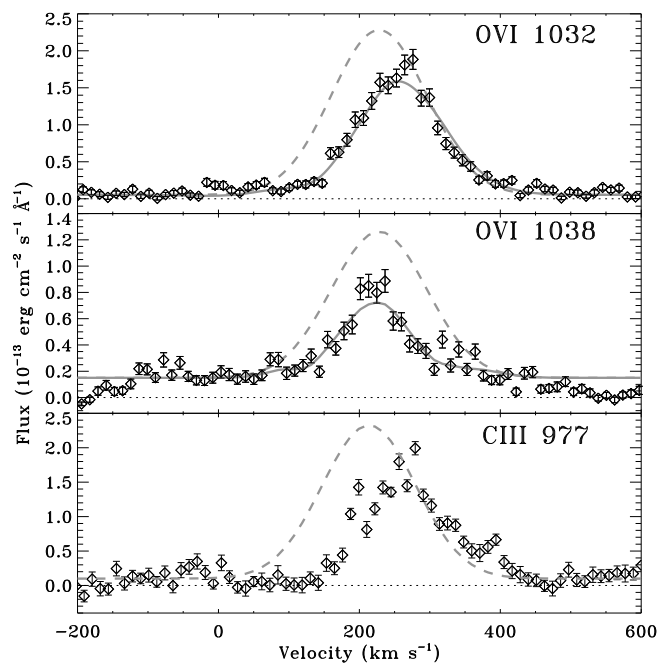}
\caption[Fits to SNR emission]
{The \OVI\ $\lambda$1032 emission (top panel) is essentially free from \HH\ absorption and is affected only on the blue side by foreground Galactic and SMC \OVI\ absorption.  The emission has been fit as described in the text (solid curve).  The intrinsic Gaussian implied by the fit is shown as the dashed curve.  We scale this intrinsic profile by a factor of two in the middle panel (dashed curve) and quantify the absorption in the more heavily contaminated \OVI\ $\lambda$1038 line.  The solid line represents the model fit.  The fits are consistent with a 2:1 optically thin ratio for the \OVI\ lines.  The bottom panel shows \CIII\ emission along with the \OVI\ $\lambda$1032 intrinsic profile for comparison.  No fitting has been attempted on the \CIII\ data.}
\end{figure}

%%% Figure 5 %%% Absorption Column
\begin{figure}
\epsscale{.75}\plotone{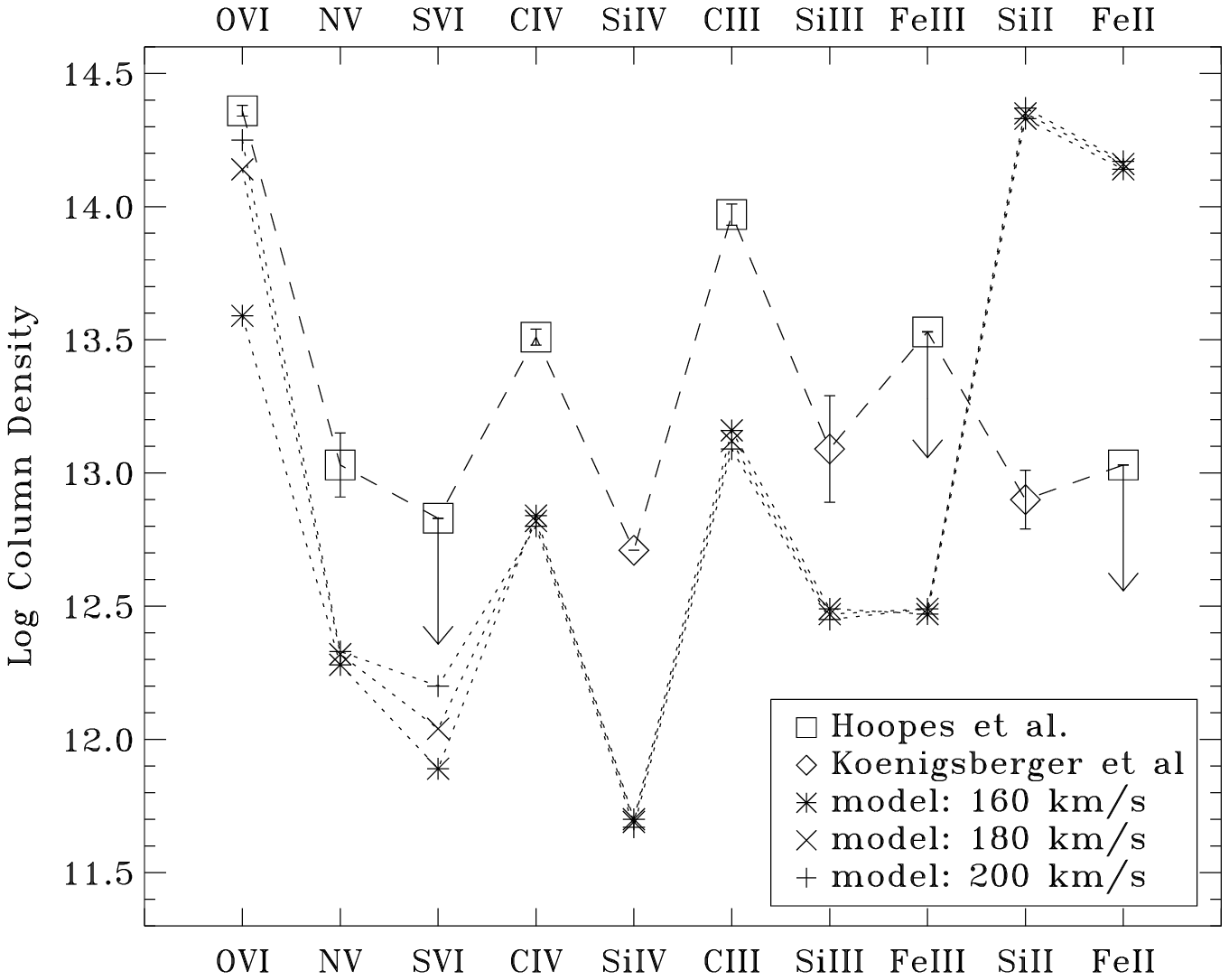}
\caption[Observed and Modeled Absorption Columns]
{Observed and modeled absorption columns for various ions toward HD\,5980.  The squares and diamonds show measured column density and quoted uncertainties from \citet{Hoopes01} (squares) and \citet{Koenigsberger01} (diamonds).  Ions are shown in order of decreasing ionization potential from \OVI\ to \FeII.  Measured columns for \ion{S}{6}, \FeIII, and \FeII\ are upper limits.  The dotted lines show the columns predicted by shock models.  In all three models, SMC abundance and preshock density $\rm n_o=1$ cm$^{-3}$ are used.  The observed columns are offset from the 160 \kms\ model by $0.78\pm0.14$ dex.  Since the model columns scale linearly with preshock density, this gives $\rm n_o=6\pm2$ cm$^{-3}$.}
\end{figure}

%%% Figure 6 %%% SNR schematic
\begin{figure}
\epsscale{.5}\plotone{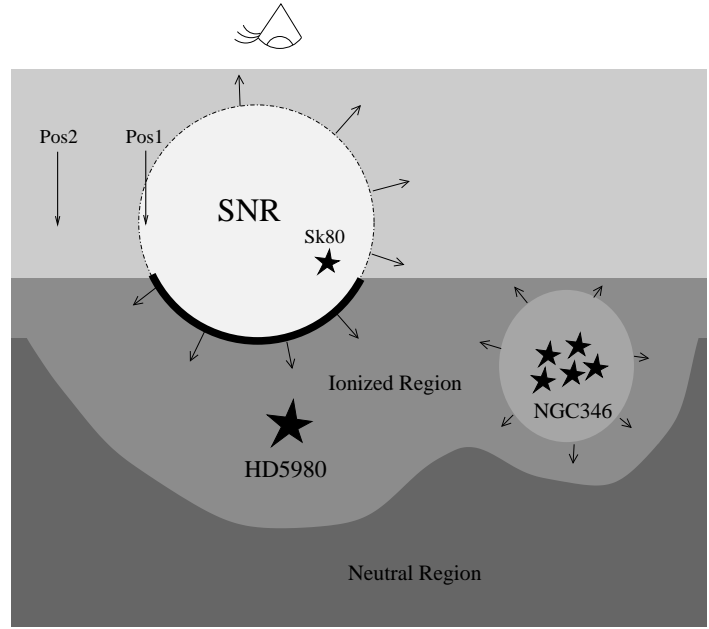}
\caption[Interaction of N66 and \SNR]
{A schematic representation of the proposed physical relationship between \SNR\ and N66 is shown.  Our line of sight is assumed to be from the top of the Figure.  \SNR\ is located on the near side of N66 and is interacting with the denser, ionized gas on the rear side.  The \OVI, \CIII, and X-ray emission seen arises from this shock interaction.  The approaching side of the shock is propagating through a more diffuse medium and is harder to detect.  HD\,5980 is located behind the SNR while Sk80 is either within or in front of the SNR.  The massive stars in the field have ionized a layer of nebular material and keep the swept-up gas behind the shockwave from recombining completely.  A slowly expanding bubble around NGC\,346 is powered by stellar winds and thermal pressure.}
\end{figure}

%%% Figure 7 %%% Frontside OVI absorption
\begin{figure}
\epsscale{.5}\plotone{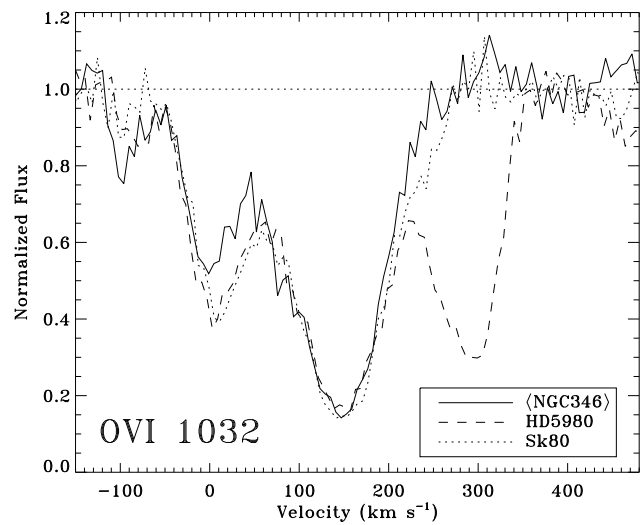}
\caption[Blue-shifted \OVI\ absorption from \SNR]
{The normalized \OVI\ absorption from HD\,5980 (dashed) and Sk80 (dotted) are markedly different from the mean NGC\,346 data (solid) near v=0 \kms.  We propose that this excess absorption represents the approaching (blue-shifted) side of the SNR.  Absorption from the receding side is clearly seen in the spectrum of HD\,5980 near 300 \kms, as reported by \citet{Hoopes01} but is not evident in the other spectra.}  
\end{figure}

\end{document}